\documentclass[aps,nofootinbib,longbibliography,prd,twocolumn]{revtex4-1}
\usepackage[babel]{csquotes}
\usepackage{graphicx}
\usepackage{amsmath,amssymb}
\usepackage[colorlinks,citecolor=blue,linkcolor=blue,urlcolor=blue]{hyperref}
\usepackage{mathrsfs}

\begin{document}
\title{``Space is blue and birds fly through it"}
\author{Carlo Rovelli}
\address{CPT, Aix-Marseille Universit\'e, Universit\'e de Toulon, CNRS, F-13288 Marseille, France.}

\begin{abstract}
\noindent Quantum mechanics is not about `quantum states': it is about values of physical variables.  I give a short fresh presentation and update on the \emph{relational} perspective on the theory, and a comment on its philosophical implications.  \\{\em [Presented to the meeting ``Foundations of quantum mechanics and their impact on contemporary society", The Royal Society, London, 11--12/12/2017; to appear in Philosophical Transactions A.]}
\end{abstract}
\maketitle

\section{A misleading notion: quantum state}

In his  celebrated 1926 paper \cite{Schrodinger:1926fk}, Erwin Schr\"odinger introduced the wave function $\psi$ and computed the Hydrogen spectrum  from first principles.   

But the theory we call ``quantum mechanics" (QM) was born one year earlier, in the work of Werner Heisenberg \cite{Heisenberg1925a}, and had already evolved into its current \emph{full} set of equations in a spectacular series of articles by Born, Jordan and Heisenberg \cite{Born1925b,Born1926}.  Dirac, following Heisenberg's breakthrough, got to the same structure independently, in 1925, the year before Schr\"odinger's work, in a work titled ``The fundamental equations of quantum mechanics" \cite{Dirac1925}. (See \cite{Fedak2009,Waerden1967} for a historical account.) Even the Hydrogen spectrum had been computed by Pauli in \cite{Pauli1926}, using the language of Heisenberg, Born and Jordan, based on  the equations  
\begin{align}\label{1.2}
\begin{split}
[q,p] =i\hbar, & \ \ \ \ \ \ \  \frac{dA}{dt} =-\frac{i}{\hbar}\,[A,H] 
\end{split}
\end{align}
and the relation between physical values and eigenvalues, with no reference to $\psi$. 

So, what did Schr\"odinger do, in his 1926 paper?  

With hindsight, he took a \emph{technical}  and a  \emph{conceptual} step. The \emph{technical} step was to change the  algebraic language of the theory, unfamiliar at the time, into a familiar one: differential equations. This brought ethereal quantum theory down to the level of the average theoretical physicist.  

The \emph{conceptual} step was to introduce the notion of ``wave function" $\psi$, soon to be evolved into the notion of ``quantum state" $\psi$, endowing it with heavy ontological weight.    This conceptual step was wrong, and dramatically misleading. We are still paying the price for the confusion it has generated.  

The confusion got into full shape in the influential second paper of the series \cite{Schrodinger1926}, where Schr\"odinger stressed the analogy with optics: the trajectory of a particle is like the trajectory of a light ray: an approximation for the behaviour of an underlying physical wave in physical space. That is:  $\psi$ is the ``actual stuff", like the electromagnetic field is the ``actual stuff" underlying the nature of light rays.  

Notice that this step is entirely ``interpretational".  It doesn't add anything to the predictive power of the theory, because this was already fully in place in the previous work of Heisenberg, Born and Jordan, where the  ``quantum state" does not play such a heavy role.  Schr\"odinger's conceptual step provided only a (misleading) way of reconceptualising the theory. 

The idea that the quantum state $\psi$ represents the ``actual stuff" described by quantum mechanics has pervaded later thinking about the theory. This is largely due to the toxic habit of introducing students to quantum theory beginning with Schr\"odinger's ``wave mechanics": thus betraying at the same time  history,  logic, and reasonableness. 

The founders of quantum mechanics saw immediately the mistakes in this step. Heisenberg was vocal in pointing them out \cite{Kumar2008}. 
First, Schr\"odinger's basis for giving ontological weight to $\psi$ was the claim that quantum theory is a theory of waves {\em in physical space}. But this is wrong: already the quantum state of two particles cannot be expressed as a collection of functions on physical space. Second, the wave formulation misses the key feature of atomic theory: energy discreteness, which must be recovered by additional ad hoc assumptions, since there is no reason for a physical wave to have energy related to frequency.  Third, and most importantly, if we treat the ``wave" as the real stuff, we fall immediately into the horrendous ``measurement" problem.  In its most vivid form (due to Einstein): how can a ``wave", spread over a large region of space, suddenly concentrate on a single place where the quantum particle manifests itself?   

All these obvious difficulties, which render the ontologicisation of $\psi$ absurd, were rapidly pointed out by Heisenberg. But Heisenberg lost the political battle against Schr\"odinger, for a number of reasons.  First, all this was about ``interpretation" and for many physicists this wasn't so interesting after all, once the equations of quantum mechanics begun producing wonders.  Differential equations are easier to work with and sort of visualise, than non-commutative algebras.  Third, Dirac himself, who did a lot directly with non-commutative algebras, found it easier to make the calculus concrete by giving it a linear representation on Hilbert spaces, and von Neumann followed: on the one hand, his robust mathematical formulation of the theory brilliantly focused on the proper relevant notion: the non-commutative observable algebra, on the other, the weight given to the Hilbert space could be taken by some as an indirect confirmation of the ontological weight of the quantum states.  Fourth, and most importantly, Bohr ---the recognised fatherly figure of the  community--- tried to mediate between his two brilliant bickering children, by obscurely agitating hands about a shamanic ``wave/particle duality".   To be fair, Schr\"odinger himself realized soon the problems with his early interpretation, and became one of the most insightful contributors to the long debate on the interpretation; but the misleading idea of taking the ``quantum state'' as a faithful description of reality stuck.   

If we want to get any clarity about quantum mechanics what we need is to undo the conceptual confusion raised by Schr\"odinger's introduction of the quantum state $\psi$. 

The abstract of the breakthrough paper by Heisenberg reads:  ``The aim of this work is to set the basis for a theory of quantum mechanics based exclusively on relations between quantities that are in principle observable."  Only relations between variables, not new entities.  The philosophy is not to inflate ontology: it is to rarefy it. 

Felix Bloch reports an enlightening conversation with Heisenberg \cite{Bloch1976}:  ``We were on a walk and somehow began to talk about space. I had just read Weyl's book Space, Time and Matter, and under its influence was proud to declare that space was simply the field of linear operations. `Nonsense,' said Heisenberg, `\emph{space is blue and birds fly through it}.'  This may sound naive, but I knew him well enough by that time to fully understand the rebuke. What he meant was that it was dangerous for a physicist to describe Nature in terms of idealised abstractions too far removed from the evidence of actual observation. In fact, it was just by avoiding this danger in the previous description of atomic phenomena that he was able to arrive at his great creation of quantum mechanics. In celebrating the fiftieth anniversary of this achievement, we are vastly indebted to the men who brought it about: not only for having provided us with a most powerful tool but also, and even more significant, for a deeper insight into our conception of reality."  

What is thus this ``deeper insight into our conception of reality", that allowed Heisenberg to find the equations of quantum mechanics, and that has no  major use of the quantum state $\psi$? 

\section{Quantum theory as a theory of physical variables}

Classical mechanics describes the world in terms of physical variables.  Variables take values, and these values describe the events of nature. Physical systems are characterised by sets of variables and interact. In the interaction, system affect one another in a manner that depends on the value taken by their variables.  Given knowledge of some of these values, we can, to some extent, predict more of them. 

The same does quantum mechanics.  It describes the world in terms of physical variables.  Variables take values, and these values describe the events of nature.  Physical systems are characterised by sets of variables and interact, affecting one another in a manner that depends on the value taken by their variables. Given knowledge of some values, we can, to some extent, predict more of them. 

The basic structure of the two theories is therefore the same.  The differences between classical and quantum mechanics are three, interdependent:
\begin{enumerate}
\item[(a)] There is fundamental \emph{\bf discreteness} in nature, because of which many physical variables can take only certain specific values and not others.
\item[(b)] Predictions can be made only  \emph{\bf probabilistically}, in general. 
\item[(c)] The values that a variables of a physical system takes are such only \emph{\bf relative} to another physical system.  Values taken relatively to distinct physical systems do not need to precisely fit together coherently, in general.  
\end{enumerate}
I discuss with more precision these three key aspects of quantum theory, from which all the rest follows, below.  The first --discreteness-- is the most important and characteristic: it gives the theory its name. It is curiously disregarded in many, if not most, philosophers' discussions on quantum theory. The third is the one with heavy philosophical implications, which I shall briefly touch below.  

This account of the theory is the interpretative framework called ``Relational QM". It was introduced in 1996 in \cite{Rovelli:1995fv} (see also \cite{Laudisa2017,Laudisa2001,Smerlak:2006gi,Rovelli2016}). In the philosophical literature it as been extensively discussed by Bas van Fraassen  \cite{Fraassen:2010fk} from a marked empiricist perspective, by Michel Bitbol \cite{pittphilsci3506,Bitbol2010} who has given a neo-Kantian version of the interpretation, by Mauro Dorato \cite{Dorato2013a} who has defended it against a number of potential objections and discussed its philosophical implication on monism, and recently by Laura Candiotto \cite{Candiotto2017} who has given it an intriguing reading in terms of (Ontic) Structural Realism.  Metaphysical and epistemological implications of relational QM have also been discussed by  Matthew Brown \cite{Brown2009a} and Daniel Wolf (n\'e Wood) \cite{Wood2010}.

\subsection{Discreteness}

I find it extraordinary that so many philosophical discussions ignore the main feature of quantum theory: discreteness.   

Discreteness is not an accessory  consequence of quantum theory, it is its core.  Quantum theory is characterised by a physical constant: the Planck constant $h=2\pi\hbar$. This constant sets the scale of the discreteness of quantum theory, and thus determines how bad is the approximation provided by classical mechanics.  Several ``interpretations" of quantum theory seem to forget the existence of the Planck constant and offer no account of its physical meaning.  

Here is a more detailed account of discreteness:  

A physical system is characterised by an ensemble of variables. The space of the possible values of these variables is the \emph{phase space} of the system.  For a system with a single degree of freedom, the phase space is two dimensional. Classical physics assumes that the variables characterising a system have always a precise value, determining a point in phase space. Concretely we never determine a point in phase space with infinite precision ---this would be meaningless---, we rather say that the system ``is in a finite region $R$ of phase space", implying that determining the value of the variables will yield values in $R$. Classical mechanics assumes that the region $R$ can be arbitrarily small. 

Now, the volume $Vol(R)$ of a region $R$ of phase space has dimensions $Length^2\times Mass/Time$, for each degree of freedom.  This combination of dimensions, $Length^2\times Mass/Time$, is called `action'  and is the  dimension of the Planck constant.  Therefore what the Planck constant fixes is the size of a (tiny) region in the space of the possible values that the variables of \emph{any} system can take. 

Now: \emph{the} major physical characterisation of quantum theory is that the volume of the region $R$ where the system happens to be cannot be smaller that $2\pi \hbar$:
\begin{align}\label{1.1}
\begin{split}
Vol(R) &\ge 2\pi\hbar
\end{split}
\end{align}
per each degree of freedom.  This is the most general and most important physical fact at the core of quantum theory. 

This implies that the number of possible values that \emph{any} variable distinguishing points within the region $R$ of phase space and which can be determined without altering the fact that the system is in the region $R$ itself, is at most
\begin{align}\label{1.3}
\begin{split}
N\le\frac{Vol(R)}{2\pi\hbar}
\end{split}
\end{align}
which is a \emph{finite} number. That is, this variable can take \emph{discrete} values only. If it wasn't so, the value of the variable could distinguish arbitrary small regions of phase space, contradicting \eqref{1.1}.  In particular: any variable separating finite regions of phase space is necessarily discrete.  

Quantum mechanics provides a precise way of coding the possible values that a physical quantity can take.  Technically: variables of a system are represented by  (self-adjoint) elements $A$ of a ($C^*$) algebra $\cal A$.  The values that the variable $a$ can take are the spectral values of the corresponding algebra element $A\in{\cal A}$.

\subsection{Probability}

Mechanics predicts the values of some variables, given some information on the values that another set of variables has taken.  In quantum mechanics, the available information is coded as a (normalized positive linear) functional $\rho$ over $\cal A$. This is called a `state'. The theory states that the \emph{statistical} mean value of a variable $A$ is $\rho(A)$.  Thus values of variables can be, in general, predicted only probabilistically. 

In turn, the state $\rho$ is computed from values that variables take.  (Technically: using the notation $\rho(A)={\rm Tr}[\rho A]$, a variable $b$ taking value in the interval $I$ of its spectrum, determines the state $\rho=cP^{b}_I$ where $P^b_I$ is the projector associated to $I$ in the spectral resolution of $B$ and $c$ is the normalization constant fixed by $\rho(1\!\!1)=1$.  If then a variable $b'$ takes value in $I'$, $\rho$ changes to $ \rho'=cP^{b'}_{I'}P^{b}_IP^{b'}_{I'}$ and so on.)  

The value of a quantity is sharp when the probability distribution is concentrated on it ($\rho(A^2)=(\rho(A))^2$). For a non-commutative  quantum algebra, there are no states where all variables are sharp. Therefore the values of the variables can never determine a point in phase space sharply.  This is the core of quantum theory, which is therefore determined by the non-commutativity of the algebra. The Planck constant $\hbar$ is the dimensional constant on the right hand side of the commutator: it determines the amount of non commutativity, hence discreteness, hence impossibility of sharpness of all variables.  

The non-commutativity between variables is Heisenberg breakthrough, understood and formalised by Born and Jordan, who were the first to write the celebrated relation $[q,p]=i\hbar$ and to recognize this non-commutativity as the key of the new theory, in 1925. 

The non-commutativity of the algebra of the variables (measured by $\hbar$) is the mathematical expression of the physical fact that variables cannot be  simultaneously sharp, hence there is a ($\hbar$-size) minimal volume attainable in phase space, hence predictions are probabilistic. 

The fact that values of variables can be predicted only probabilistically raises the key interpretational question of quantum mechanics: when and how is a probabilistic prediction resolved into an actual value? 

\subsection{The relational aspect of quantum theory}

When and how a probabilistic prediction about the value of a variable $a$ of a physical system $S$ is resolved into an actual value?

The answer is: when $S$ interacts with another physical system $S'$.  Value actualisation happens at interactions since variables represent the ways systems affect one another.  Any interaction counts, irrespectively of  size, number of degrees of freedom, presence of records, consciousness, degree of classicality of $S'$, decoherence, or else, because none of these pertain to elementary physics. 

In the course of the interaction, the system $S$ affects the system $S'$.  If the effect of the interaction on $S'$ depends on the variable $a$ of $S$, then the probabilistic spread of $a$ is resolved into an actual value, or, more generally, into an interval $I$ of values in its spectrum.

Now we come to the crucial point.  The actualisation of the value of $a$ is such only relative to the system $S'$.  The corresponding state $\rho'$ determined by the actualisation is therefore a state relative to $S'$, in the sense that it predicts \emph{only} the probability distribution of variables of $S$ in subsequent interactions with $S'$. {\em It has no bearing on subsequent interactions with other physical systems.}   

This is the profoundly novel relational aspect of quantum mechanics. 

Why are we forced to this extreme conclusion?  The argument, detailed in \cite{Rovelli:1995fv}, can be summarised as follows.   

We must assume that variables do take value, because the description of the world we employ is in terms of values of variables. However, the predictions of quantum mechanics are incompatible with \emph{all} variables having simultaneously a determined value. A number of mathematical results, such as the Kochen-Specker \cite{Kochen1968} theorem, confirm that if all variables could have a value simultaneously, the predictions of quantum mechanics would be violated.   Therefore, something must determine \emph{when} a variable has a value.  

The textbook answer is ``when we measure it". This obviously makes no sense, because the grammar of Nature certainly does not care whether you or I are ``measuring" anything.   Measurement is an interaction like any other.  Variables take value at any interaction. 

But, (this is the key point) if a system $S$ interacts with a system $S'$, QM predicts that in a later interaction with a further system $S''$, a variable $b$ of the $S\cup S'$ system is \emph{not} determined by $\rho'$. Rather, it is determined by the joint  dynamical evolution of the $S\cup S'$ \emph{quantum} system.  In physicists' parlance: quantum theory predicts interference effects between the different branches corresponding to different values of the variable $a$, as if no actualisation had happened.  

We have thus to combine the presence of these interference effects (which pushes us to say that $a$ had no value) with the fact that the variable $a$ does take a value.\footnote{ In Many World interpretations, $a$ takes a value indexically relative to a world; in Bohm-like theories only an (abelian) subset of variables has value, not all of them; in Quantum Information interpretations, $a$ takes a value only when the interaction is with the idealistic holder of the information; in Copenhagen-like interpretations, when the interaction is with the ``classical world"; in Physical Collapse theories, when some not yet directly observed random physical phenomenon happens...} 

The answer of relational QM is that the variable $a$ of the system $S$ actualized in the interaction with $S'$  takes value \emph{with respect to} $S'$, but not \emph{with respect to} $S''$.   This is the core idea underlying the ``relational" interpretation of quantum mechanics.   

Relationality is no surprise in physics.  In classical mechanics the velocity of an object has no meaning by itself: it is only defined with respect to another object.   The color of a quark in  strong-interaction theory has no meaning by itself: only the relative color of two quarks has meaning.   In electromagnetism, the potential at a point has no meaning, unless another point is taken as reference; that is, only relative potentials have meanings. In general relativity, the location of something is only defined with respect to the gravitational field, or with respect to other physical entities; and so on.   But quantum theory takes this ubiquitous relationalism, to a new level: the actual value of \emph{all} physical quantities of \emph{any} system is only meaningful in relation to another system.   Value actualisation is a relational notion like velocity.

\section{What is the quantum state?}

The above discussion shows that the quantum state $\rho$ does not pertain solely to the system $S$.  It pertains also to the system $S'$, because it depends on variables' values, which pertain only to $S'$.  The idea that states in quantum mechanics are relative states, namely states of a physical system relative to a second physical system is Everett's lasting contribution to the understanding of quantum theory \cite{Everett:1957hd}.   

A moment of reflection shows that the quantum states used in real laboratories  where scientists use quantum mechanics concretely is obviously \emph{always} a \emph{relative}  state.  Even a radical believer in a universal quantum state would concede that the $\psi$ that physicists use in their laboratories to describe a quantum system is not the hypothetical universal wave function: it is the relative state, in the sense of Everett, that describes the properties of the system, relative to the apparata it is interacting with. 

What precisely is the quantum state of $S$ relative to $S'$?  What is $\psi$ (or $\rho$)? The discussion above clarifies this delicate point: it is a theoretical  devise we use for bookkeeping information about the values of variables of $S$ actualised in interactions with $S'$, values which can in principle be used for predicting other (for instance future, or past) values that variables may take in other interactions with $S'$.    

Charging $\psi$ with ontological weight is therefore like charging with ontological weight a distribution function of statistical physics, or the information I have about current political events: a mistake that generates mysterious ``collapses" anytime there is an interaction.  More specifically, in the semiclassical approximation $\psi\sim e^{iS}$ where $S$ is a Hamilton-Jacobi function. This shows that the physical nature of $\psi$ is the same as the physical nature of a Hamilton-Jacobi function. Nobody in their right mind would charge $S$ with ontological weight, in the context of classical mechanics: $S$ is a calculational device used to predict an outcome on the basis of an input. It ``jumps" at each update of the calculation. 

Quantum mechanics is thus not a theory of the dynamics of a mysterious $\psi$ entity, from which mysteriously the world of our experience emerges.  It is a theory  of the possible values that conventional physical variables take at interactions, and the transition probabilities that determine which values are likely to be realized, given that others are \cite{Groenewold1957}. 

The fact that the quantum state is a bookkeeping devise that cannot be charged with ontological weight is emphasized by the following observation  \cite{Rovelli2016}.  Say I know that at time $t$ a particle interacts with a $x$-oriented Stern-Gerlach device. Then I can predict that (if nothing else  happens in between) the particle has probability $\frac12$ to be up (or down) spinning, when interacting with a $z$-oriented Stern--Gerlach devise at time $t'$.  Key point: this is true irrespectively of which between $t$ and $t'$ comes earlier.   Quantum probabilistic predictions are the same both forth and back in time.  So: what is the state of the particle in the time interval between $t$ and $t'$?  Answer: it depends only on what I know: if I know the past (respectively, future) value, I use the state to book-keep the future (respectively, past) value. The state is a coding of the value of the $x$ spin that allows me to predict the $z$ spin, not something that the particle ``has".  We can be realist about the two values of the spin, not about the $\psi$ in between, because $\psi$ depends on a time orientation, while the relevant physics does not.   

A coherent ontology for quantum mechanics is thus sparser than the classical one, not heavier.   

A good name for the actualisation of the value of a variable in an interaction is ``quantum event".  The proper ontology for quantum mechanics is a sparse ontology of (relational) quantum events happening at interactions between physical systems. 

\subsection{Information}

Equation \eqref{1.1} can be expressed by saying that 

(P1) {\em The amount of information that can be extracted from a finite region of phase space is finite.} 

``Information" means here nothing else than ``number of possible distinct alternatives".   

The step from $\rho$ to $\rho'$ determined by an actualisation modifies the predictions of the theory.  In particular, the value of $a$, previously spread, is then predicted to be sharper. This can be expressed in information theoretical theorems by saying that 

(P2) {\em An interaction allows new information about a system to be acquired.} 

There is an apparent tension between the two statements (P1) and (P2). If there is a finite amount of information, how can we keep gathering novel information?  The tension is only apparent, because here `information' quantifies the data relevant for predicting the value of variables. In the course of an interaction,  part of the previously relevant information becomes irrelevant.  In this way, information is acquired, but the total amount of information available remains finite.\footnote{ Here is a simple example: if a spin-$\frac12$ particle passes through a $z$-oriented Stern-Gerlach apparatus and takes the ``up" path, we have one bit of information about the orientation of its angular momentum $\vec L$.  If it then crosses an $x$-oriented apparatus, we gain one bit of information (about $L_x$) and we loose one bit of information (about $L_z$).} 

It is the combination of (P1) and (P2) that largely characterises quantum theory (for the case of qubit-systems, see \cite{Hoehn2014}). These two statements were proposed as the basic ``postulates" of quantum mechanics in \cite{Rovelli:1995fv}. The apparent contradiction between the two capturing the counterintuitive character of QM in the same sense in which the apparent contradiction between the two Einstein's postulate for Special Relativity captures the counterintuitive character of relativistic spacetime geometry. Very similar ideas were independently introduced by Zeilinger and Brukner \cite{Zeilinger:1999bh,Brukner:2003dq}. 

An attempt to reconstruct the full formalism of quantum theory starting from these two information-theoretic postulated was initiated in \cite{Rovelli:1995fv} (see also \cite{Grinbaum2005}). Recently a remarkable reconstruction theorem along these lines has been successfully completed for the case of finite dimensional systems in \cite{Hoehn2014,Hohn2017}, shedding considerable new light on the structure of the theory and its physical roots. 

The role of information at the basis of quantum theory is a controversial topic.  The term `information' is ambiguous, with a wide spectrum of meanings ranging from epistemic states of conscious observers all the way to simply counting alternatives, \`a la Shannon. As pointed out for instance by Dorato, even in its weakest sense information cannot be taken as a primary notion from which all others can be derived, since it is always information about something.  Nevertheless, information can be a powerful organisational principle in the sense of Einstein's distinction between `principle theories' (like thermodynamics) versus `constructive theories' (like electromagnetism) \cite{Spekkens2014}. The role of the general theory of mechanics is not to list the ingredients of the world ---this is done by the individual mechanical theories, like the Standard Model of particle physics, general relativity, of the harmonic oscillator.  The role of the general theory of mechanics (like classical mechanics or quantum mechanics) is to provide a general framework within which specific constructive theories are realized.  From this perspective, the notion of information as number of possible alternatives may play a very useful role. 

It is in this sense that the two postulates can be understood. They are limitations on the structure of the values that variables can take. The list of relevant variables, which define a physical system, and their algebraic relations, are provided by specific quantum theories. 

There are several objections that come naturally to mind when one first encounters relational QM, which seem to render it inconsistent.  These have been long discussed and have all been convincingly answered, see in particular the detailed arguments in van Fraassen \cite{Fraassen:2010fk} and Dorato \cite{Dorato2013a} and the original paper \cite{Rovelli:1995fv}; I will not re-discuss them here. Relational QM is a consistent interpretation of quantum theory. 

But, like all other consistent interpretations, it comes at a price. 

%
%

\section{Philosophical implications}

\subsection{Every interpretation has a cost}

Every interpretation of quantum theory comes with a ``cost".  

Examples from some interpretations popular nowadays are the following.  The cost of the Many World interpretations is the hugely inflated ontology of a continuous family of equally existing ``worlds", of which we basically know nothing, and an awkward difficulty of rigorously recovering the actual values of the variables in terms of which we describe the world, from the pure-$\psi$ picture taken as fundamental.  The cost of the Physical Collapse interpretations, such as the Ghirardi-Weber-Rimini theory, is to be based on physics which is so far unobserved and that many physicists view as not very plausible.  The cost of the Bohmian interpretations is to postulate the existence of stuff which is unobservable \emph{in principle} and which, in th eview of most physicists, violates too badly what we have learned about Nature in the last century.  The cost of Quantum Informational interpretations (partially inspired by relational QM \cite{Fuchs2001}) is to be tied to a basically idealistic stance where the holder of the information is treated as an unphysical entity, a priori differently from all other physical systems, which cannot be in superpositions. The so called Copenhagen Interpretations, which are held by the majority of real physicists concretely working with quantum mechanics,  postulate the existence of a generally ill-explained ``classical world", whose interactions collapse quantum states. And so on... 

Do not take these criticisms badly: they are not meant to dismiss these interpretations, they are simply the reasons commonly expressed for which each interpretation does not sound plausible to others: the point I am making is that there is no interpretation of quantum mechanics that is not burdened by a heavy cost of some sort, which appears too heavy a price to pay to those who do not share the passion for that particular interpretation.  Many discussions about quantum theory are just finger-pointing to one another's cost.  

The evaluation of these costs depends on wider philosophical perspectives that we explicitly or implicitly hold.  Attachment to full fledged strong realism leads away from Quantum Informational interpretations and towards Bohm or Many Worlds. Sympathy for empiricism or idealism leads in opposite directions, towards Copenhagen or Quantum Information. And so on; the picture could be fine-grained. 

The beauty of the problem of the interpretation of quantum mechanics is precisely the fact that the spectacular and unmatched empirical success of the theory forces us to give up at least some cherished philosophical assumption.   Which one is convenient to give up is the open question. 

The relational interpretation does not escape this dire situation.  As seen from the reactions in the philosophical literature, relational QM is compatible with  diverse philosophical perspectives. But not all.  How strong is the philosophical ``cost" of relational QM? 

Its main cost is a challenge to a strong version of realism, which is implied by its radical relational stance.  

\subsection{Realism}

`Realism' is a term used with different meanings.  Its weak meaning is the assumption that there is a world outside our mind, which exists independently from our perceptions, beliefs or thoughts.  

Relational QM is compatible with realism in this weak sense.  ``Out there" there are plenty of physical systems interacting among themselves and about which we can get reliable knowledge by interacting with them; there are plenty of variables taking values, and so on. There is nothing similar to `mind' required to make sense of the theory.  What is meant by a variable taking value ``with respect to a system $S'$" is not $S'$ to be a conscious subject of perceptions ---it just the same as when we say that the velocity of the Earth is $40km/s$ ``with respect to the sun": no implication of the sun being a sentient being ``perceiving" the Earth.   In this respect, quantum theory is no more and no less compatible with realism (or other metaphysics) than classical mechanics. I myself think that we, conscious critters, are physical systems like any other. Relational QM is anti-realist about the wave function, but is realist about quantum events, systems, interactions...  It maintains that ``space is blue and birds fly through it" and space and birds can be constituted by molecules, particles, fields, or whatever. What it denies is the utility --even the coherence-- of thinking that all this is made up by some underlying $\psi$ entity.   

But there is a stronger meaning of `realism': to assume that it is in principle possible to list all the features of the world, all the values of all variables describing it at some fundamental level, at each moment of continuous time, as is the case in classical mechanics. This is not possible in relational QM.   Interpretations of QM that adhere to strong realism, like Many Worlds, or Bohm, or other hidden variables theories, circumvent the Kochen-Specker theorem, which states that in general there is no consistent assignment of a definite values to all variables, by restricting the set of elementary variables describing the world (to the quantum state itself, or to Bohmian trajectories, or else).  Relational QM assumes seriously the Kochen-Specker theorem: variables take value only at interactions.  The stronger version of the realist credo is therefore in tension with QM, and this is at the core of relational QM.   It is not even realized in the relatively weaker sense of considering a juxtaposition of all possible values relative to all possible systems. The reason is that the very fact that a quantity has value with respect to some system is itself relative to that system \cite{Rovelli:1995fv}. 

This weak realism of relational QM is in fact quite common in physics laboratories.  Most physicists would declare themselves ``realists", but not realists about $\psi$. As one of the two (very good) referees of this paper put it: ``In physicistsÕ circles, Schr\"odingerÕs $\psi$ is mostly regarded as a mere instrument".  Relational QM is a way to make this position coherent. 

There are three specific challenges to strong realism that are implicit in relational QM.  

The first is its sparse ontology.  The question of ``what happens between quantum events" is meangless in the theory.  The happening of the world is a very fine-graned but discrete swarming of quantum events, not the permanence of entities that have well defined properties at each moment of a continuous time. 

This is the way the relational interpretation circumvents results like the Pusey-Barrett-Rudolph theorem \cite{Pusey}. Such theorems \emph{assume} that at every moment of time all properties are well defined. (For a review, see \cite{Leifer2014}) They essentially say that if there is a hidden variable theory, the hidden variables must contain at least the entire information which is in the quantum state.  But the assumption is explicitly denied in relational QM: properties do not exist at all times: they are properties of events and the events happen at interactions. 

On the same vein, in \cite{Laudisa2017a} Laudisa criticises relational QM because it does not provide a ``deeper justification" for the ``state reduction process".  This is like criticising classical mechanics because it it does not provide a ``deeper justification" for why a system follows its equations of motion.  It is a stance based on a very strong realist (in the narrow sense specified above) philosophical assumption.   In the history of physics much progress has happened by realising that some naively realist expectation was ill founded, and therefore by dropping these kind of questions: How are the spheres governing the orbits of planet arranged? What is the mechanical underpinning of the electric and magnetic fields? Into where is the universe expanding? To some extent, one can say that modern science itself was born in Newton's celebrated ``hypotheses non fingo", which is precisely the recognition that questions of this sort can be misleading.  When everybody else was trying to find dynamical laws accounting for atoms, Heisenberg's breakthrough was to realise that the known laws where already good enough, but the ontology was sparser and the question of the actual continuous orbit of the electron was ill posed.  I think that we should not keep asking what amounts to this same question over and over again: trying to fill-in the sparse ontology of Nature with our classical intuition about continuity. On this, see the enlightening discussion given by Dorato in \cite{Dorato2017}.

The second element of relational QM that challenges a strong version of realism is that values taken with respect to different systems can be compared \cite{Rovelli:1995fv} (hence there no solipsism), but the comparison amounts to a physical interaction, and its sharpness is therefore limited by $\hbar$. Therefore we cannot escape from the limitation to partial views: there is no coherent global view available.   Matthew Brown has discussed this point in \cite{Brown2009a}. 

The third, emphasized by Dorato, is the related `anti-monistic' stance implicit in relational QM. Since the state of a system is a bookkeeping device of interactions with \emph{something else}, it follows immediately that there is no meaning in ``the quantum state of the full universe". There is no something else to the universe. Everett's relative states are the only quantum states we can meaningfully talk about. Every quantum state is an Everett's quantum state.  A reason for rejecting relational QM, therefore, comes if we assume that the monistic idea of the ``state of the whole" must makes sense and must be  coherently given in principle.\footnote{ This does not prevent conventional quantum cosmology to be studied, since physical cosmology is not the science of everything: it is the science of the largest-scale degrees of freedom.}  

This relational stance of relational QM requires a philosophical perspective where relations play a central role. This is why Candiotto \cite{Candiotto2017} suggest to frame relational QM in the general context of Ontic Structural Realism. This is certainly an intriguing possibility.   My sympathy for a natural philosophical home for relational QM is an anti-foundationalist perspective where we give up the notion of primary substance carrying attributes, and recognize the mutual dependence of the concepts we use to describe the world. Other perspectives are possible, as we have seen in the strictly empiricists and neo-Kantian readings by van Fraassen and Bitbol, but strong realism in the strict sense of substance and attributes that are always uniquely determined is not. 

\subsection{How to go ahead?}

There are three developments that could move us forward.  

The first is novel empirical information. Some interpretations of quantum theory lead to empirically distinguishable versions of the theory. Empirical corroboration of their predictions would change the picture; repeated failure to detect discrepancy from standard QM weakens their credibility.   This is the way progress happens in experimental physics.  So far, QM has been unquestionably winning for nearly a century, beyond all expectations.

The second is theoretical fertility.  If for instance quantum gravity turned out to be more easily comprehensible in one framework than in another, then this framework would gain credibility.  This is the way progress happens in theoretical physics.  

My focus on relational QM, indeed, is also motivated by my work in quantum gravity \cite{Rovelli:2004fk,Rovelli:2014ssa}. In quantum gravity, where we do not have a background spacetime where to locate things, relational QM works very neatly because the quantum relationalism combines in a surprisingly natural manner with the relationalism of general relativity.  Locality is what makes this work.  Here is how \cite{Vidotto2013}: the quantum mechanical notion of ``physical system" is identified with the general relativistic notion of ``spacetime region".  The quantum mechanical notion of ``interaction" between systems is identified with the general relativistic notion of ``adjacency" between spacetime regions.  Locality assures that interaction requires (and defines) adjacency.  Thus quantum states are associated to three dimensional surfaces bounding spacetime regions and  quantum mechanical transition amplitudes are associated to ``processes" identified with the spacetime regions themselves. In other words, variables actualise at three dimensional boundaries, with respect to (arbitrary) spacetime partitions.   The theory can then be used locally, without necessarily assuming anything about the global aspects of the universe.

The third manner in which progress can happen is how it does in philosophy: ideas are debated, absorbed, prove powerful, or weak, and slowly are retained or discarded.  I am personally actually confident that this can happen for quantum theory.  

The key to this, in my opinion, is to fully accept this interference between the progress of fundamental physics and some major philosophical issues, like the question of realism, the nature of entities and relations, and the question of idealism.  Accepting the reciprocal interference means in particular to reverse the way general philosophical stances color our preferences for interpretation.  That is, rather than letting our philosophical orientation determine our reading of QM, be ready to let the discoveries of fundamental physics influence our philosophical orientations. 

It woundn't certainly be the first time that philosophy is heavily affected by science.  I believe that we should not try to understand the world rigidly in terms of our conceptual structure.  Rather we should humbly allow our conceptual structure to be moulded by empirical discoveries.  This, I think, is how knowledge develops best. 

Relational QM is a radical attempt to directly cash out the initial breakthrough that originated the theory: the world is described by variables that can take values and obey the equations of classical mechanics, \emph{but} products of these variable have a tiny non-commutativity that generically prevents sharp value assignment, leading to discreteness, probability and to the relational character of the value assignment.  

The founders of the theory expressed this relational character in the ``observer-measurement" language. This language seems to require that special systems (the observer, the classical world, macroscopic objects...) escape the quantum limitations.  But neither of this, and in particular no ``subjective states of conscious observers", is needed in the interpretation of QM.   As soon as we relinquish this exception, and realize that \emph{any} physical system can play the role of a Copenhagen's ``observer", we fall into relational QM.  Relational QM is Copenhagen quantum mechanics made democratic by bringing all systems onto the same footing.


\providecommand{\href}[2]{#2}\begingroup\raggedright\endgroup

\end{document}